\documentclass[12pt,a4paper]{article}
\usepackage{graphicx}
\usepackage{amsmath}
\usepackage{amssymb}
\usepackage{epsfig}
\usepackage{times}
\begin{document}

\title{Gauge Principle and QED\footnote{Invited talk at PHOTON2005,
\textit{The Photon: Its First Hundred Years and the Future}, 31.8-04.09, 2005, Warsaw.}}
\author{
Norbert Straumann\\
Institute for Theoretical Physics\\
University of Z\"urich, Switzerland}
\date{August, 2005}
\maketitle

\begin{abstract}
\noindent One of the major developments of twentieth century physics has been the gradual
recognition that a common feature of the known fundamental interactions is their gauge
structure. In this talk the early history of gauge theory is reviewed, emphasizing
especially Weyl's seminal contributions of 1918 and 1929.
\end{abstract}

\section{Introduction}
The organizers of this conference asked me to review the early history of gauge theories.
Because of space and time limitations I shall concentrate on Weyl's seminal papers of 1918
and 1929. Important contributions by Fock, Klein and others, based on Kaluza's five-dimensional
unification attempt, will not be discussed. (For this I refer to \cite{RS} and \cite{JO}.)

The history of gauge theories begins with GR, which can be regarded as a non-Abelian gauge
theory of a special type. To a large extent the other gauge theories emerged in a slow and
complicated process gradually from GR. Their common geometrical structure -- best
expressed in terms of connections of fiber bundles -- is now widely recognized.

It all began with H. Weyl \cite{Wey2} who made in 1918 the first attempt to extend GR in
order to describe gravitation and electromagnetism within a unifying geometrical framework.
This brilliant proposal contains the germs of all mathematical aspects of non-Abelian gauge
theory. The word `gauge' (german: `Eich-') transformation appeared for the first time in
this paper, but in the everyday meaning of change of length or change of calibration.

Einstein admired Weyl's theory as ``a coup of genius of the first rate'', but immediately
realized that it was physically untenable. After a long discussion Weyl finally admitted
that his attempt was a failure as a physical theory. (For a discussion of the intense
Einstein-Weyl correspondence, see Ref. \cite{NS1}.) It paved, however, the way for the
correct understanding of gauge invariance. Weyl himself reinterpreted in 1929 his original
theory after the advent of quantum theory in a grand paper \cite{Wey3}. Weyl's
reinterpretation of his earlier speculative proposal had actually been suggested before by
London \cite{Lon}. Fock \cite{Fo}, Klein \cite{Kl1}, and others arrived at the principle of
gauge invariance in the framework of wave mechanics along a completely different line. It
was, however, Weyl who emphasized the role of gauge invariance as a \emph{constructive
principle} from which electromagnetism can be derived. This point of view became very
fruitful for our present understanding of fundamental interactions. We\footnote{Soon after
our joint paper appeared in print Lochlain O`Raifeartaigh died suddenly, to the great sorrow
and surprise of his family and numerous friends. I would like to dedicate this
contribution to the memory of Lochlain.} have described this more extensively in \cite{RS}.

\begin{figure}\label{fig1}
\begin{center}
\epsfig{file=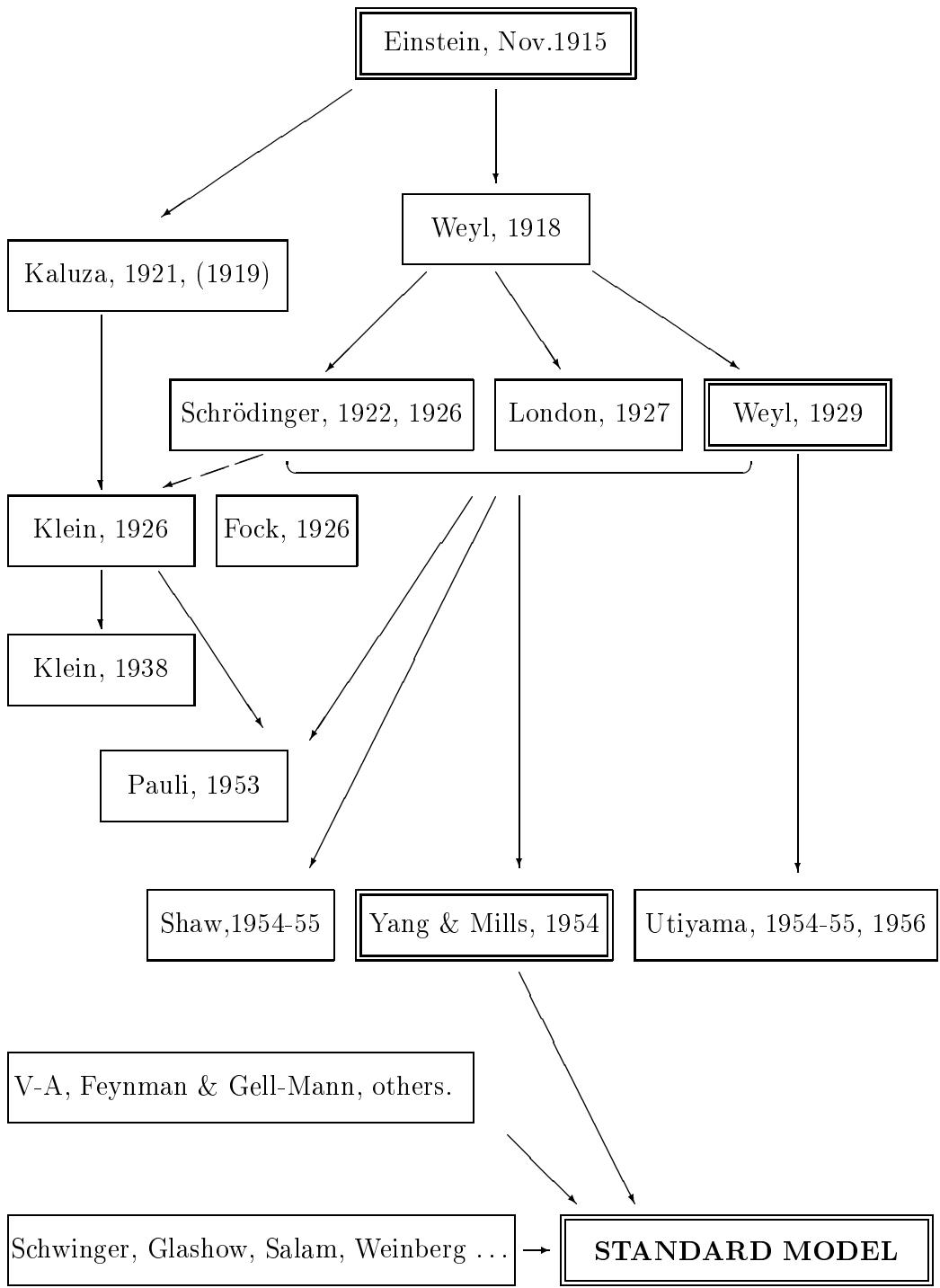,height=19cm}
\caption{Key papers in the development of gauge theories.}
\end{center}
\end{figure}

These works underlie the diagram in Fig.~\protect\ref{fig1}.


\section{Weyl's Attempt to Unify Gravitation
\protect\newline and Electromagnetism}

On the 1$^{st}$ of March 1918 Weyl writes in a letter to Einstein (\cite{CPA}, Vol. 8B,
Doc.472): ``These days I succeeded, as I believe, to derive electricity and gravitation
from a common source \ldots ''. Einstein's prompt reaction by postcard indicates already a
physical objection which he explained in detail shortly afterwards. Before we come to this
we have to describe Weyl's theory of 1918.

\subsection{Weyl's Generalization of Riemannian Geometry}

Weyl's starting point was purely mathematical. He felt a certain uneasiness about
Riemannian geometry, as is clearly expressed by the following sentences early in his paper:
\begin{quotation}
{\sl But in Riemannian geometry described above there is contained a last element of
geometry ``at a distance''  (ferngeometrisches Element) --- with no good reason, as far as
I can see; it is due only to the accidental development of Riemannian geometry from
Euclidean geometry. The metric allows the two magnitudes of two vectors to be compared, not
only at the same point, but at any arbitrarily separated points.} {\it A true infinitesimal
geometry should, however, recognize only a principle for transferring the magnitude of a
vector to an
 infinitesimally close point} {\sl and then, on transfer to an arbitrary
distant point, the integrability of the magnitude of a vector is no more to be expected
that the integrability of its direction.}
\end{quotation}

After these remarks Weyl turns to physical speculation and continues as follows:

\begin{quotation}
{\sl On the removal of this inconsistency there appears a geometry that, surprisingly, when
applied to the world,} {\it explains not only the gravitational phenomena but also the
electrical.} {\sl According to the resultant theory both spring from the same source,
indeed} {\it in general one cannot separate gravitation and electromagnetism in a unique
manner}. {\sl In this theory} {\it all physical quantities have a world geometrical
meaning; the action appears from the beginning as a pure number. It leads to an essentially
unique universal law; it even allows us to understand in a certain sense why the world is
four-dimensional}.
\end{quotation}

In brief, Weyl's geometry can be described as follows (see also ref.~\cite{AGS}). First,
the spacetime manifold $M$ is equipped with a conformal structure, i.e., with a class $[g]$
of conformally equivalent Lorentz metrics $g$ (and not a definite metric as in GR). This
corresponds to the requirement that it should only be possible to compare lengths at one
and the same world point. Second, it is assumed, as in Riemannian geometry, that there is
an affine (linear) torsion-free connection which defines a covariant derivative $\nabla$,
and respects the conformal structure. Differentially this means that for any $g\in[g]$ the
covariant derivative $\nabla g$ should be proportional to $g$:
\begin{equation} \label{2.1}
\nabla g =-2A\otimes g ~~~~~ (\nabla_{\lambda}g_{\mu\nu}=-2A_{\lambda}g_{\mu\nu}),
\end{equation}
where $A=A_{\mu}dx^{\mu}$ is a differential 1-form.

Consider now a curve $\gamma: [0,1]\rightarrow M$ and a parallel-transported vector field
$X$ along $\gamma$. If $l$ is the length of $X$, measured with a representative $g\in[g]$,
we obtain from (\ref{2.1}) the following relation between $l(p)$ for the initial point
$p=\gamma(0)$ and $l(q)$ for the end point $q=\gamma(1)$:
\begin{equation} \label{2.2}
l(q)=\exp\left(-\int_{\gamma}A\right)\ l(p).
\end{equation}
Thus, the ratio of lengths in $q$ and $p$ (measured with $g\in[g]$) {\it depends in general
on the connecting path $\gamma$} (see Fig.2). The length is only independent of $\gamma$ if
the curl of $A$,
\begin{equation} \label{2.3} F=dA ~~~~~(F_{\mu\nu}=\partial_{\mu}A_{\nu}-
\partial_{\nu}A_{\mu}),
\end{equation}
vanishes.
\begin{figure}\label{fig2}
\begin{center}
\epsfig{file=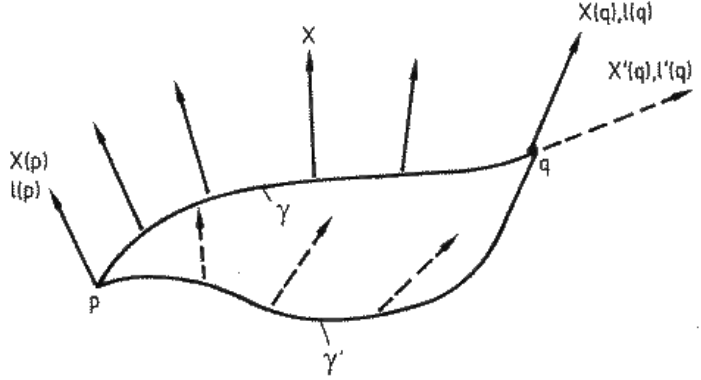,width=10cm} \caption{Path dependence of parallel displacement and
transport of length in  Weyl space.}
\end{center}
\end{figure}

The compatibility requirement (\ref{2.1}) leads to the following expression for the
Christoffel symbols in Weyl's geometry:
\begin{equation} \label{2.4}
\Gamma^{\mu}_{\nu\lambda}=\frac{1}{2}g^{\mu\sigma}(
g_{\lambda\sigma,\nu}+g_{\sigma\nu,\lambda}-g_{\nu\lambda,\sigma})
+g^{\mu\sigma}(g_{\lambda\sigma}A_{\nu}+g_{\sigma\nu}A_{\lambda}-
g_{\nu\lambda}A_{\sigma}).
\end{equation}
The second $A$-dependent term is a characteristic
new piece in Weyl's geometry which has to be added to the Christoffel symbols of Riemannian
geometry.

Until now we have chosen a fixed, but arbitrary metric in the conformal class $[g]$. This
corresponds to a choice of calibration (or gauge). Passing to another calibration with
metric $\bar{g}$, related to $g$ by
\begin{equation}\label{2.5}
\bar{g}=e^{2\lambda}g,
\end{equation}
the potential $A$ in (\ref{2.1}) will also change to $\bar{A}$, say. Since the covariant
derivative has an absolute meaning, $\bar{A}$ can easily be worked out: On the one hand we
have by definition
\begin{equation}
\nabla \bar{g} =-2\bar{A}\otimes\bar{g},
\end{equation}
and on the other hand we find for the left side with (\ref{2.1})
\begin{equation} 
\nabla\bar{g}=\nabla(e^{2\lambda}g)= 2d\lambda\otimes\bar{g}+e^{2\lambda}\nabla g=
2d\lambda\otimes\bar{g}-2A\otimes\bar{g}.
\end{equation}
Thus
\begin{equation} \label{2.6} \bar{A}=A- d\lambda
~~~(\bar{A}_{\mu}=A_{\mu}-\partial_{\mu}\lambda).
\end{equation}
This shows that a change of calibration of the metric induces a {\it ``gauge
transformation''} for $A$:
\begin{equation} \label{2.7}
g\rightarrow e^{2\lambda}g,~~~A\rightarrow A-d\lambda.
\end{equation}
Only gauge classes have an
absolute meaning. (The Weyl connection is, however, gauge-invariant. This is conceptually
clear, but can also be verified by direct calculation from expression~Eq.(\ref{2.4}).)

\subsection{Electromagnetism and Gravitation}

Turning to physics, Weyl assumes that his ``purely infinitesimal geometry'' describes the
structure of spacetime and consequently he requires that physical laws should satisfy a
double-invariance: 1. They must be invariant with respect to arbitrary smooth coordinate
transformations. 2. They must be {\it gauge invariant}, i.e., invariant with respect to
substitutions (\ref{2.7}) for an arbitrary smooth function $\lambda$.

Nothing is more natural to Weyl, than identifying $A_{\mu}$ with the vector potential and
$F_{\mu\nu}$ in eq. (\ref{2.3}) with the field strength of electromagnetism. In the absence
of electromagnetic fields ($F_{\mu\nu}=0$) the scale factor $\exp(-\int_{\gamma}A)$ in
(\ref{2.2}) for length transport becomes path independent (integrable) and one can find a
gauge such that $A_{\mu}$ vanishes for simply connected spacetime regions. In this special
case one is in the same situation as in GR.

Weyl proceeds to find an action which is generally invariant as well as gauge invariant and
which would give the coupled field equations for $g$ and $A$. We do not want to enter into
this, except for the following remark. In his first paper \cite{Wey2} Weyl proposes what we
call nowadays the Yang-Mills action
\begin{equation}\label{2.8}
S(g,A)=-\frac{1}{4}\int Tr(\Omega\wedge\ast\Omega).
\end{equation}
Here $\Omega$ denotes the curvature form and $\ast\Omega$ its Hodge dual\footnote{The
integrand in (\ref{2.8}) is in local coordinates indeed just the expression
$R_{\alpha\beta\gamma\delta} R^{\alpha\beta\gamma\delta} \sqrt{-g}dx^{0}\wedge\ldots\wedge
dx^{3}$ which is used by Weyl ($R_{\alpha\beta\gamma\delta}$= the curvature tensor of the
Weyl connection).}. Note that the latter is gauge invariant, i.e., independent of the
choice of $g\in[g]$. In Weyl's geometry the curvature form splits as
$\Omega=\hat{\Omega}+F$, where $\hat{\Omega}$ is the metric piece \cite{AGS}.
Correspondingly, the action also splits,
\begin{equation} \label{2.9}
Tr(\Omega\wedge\ast\Omega) = Tr (\hat{\Omega}\wedge\ast\hat{\Omega}) +F\wedge\ast F.
\end{equation}
The second term is just the Maxwell action. Weyl's theory thus contains formally
all aspects of a non-Abelian gauge theory.

Weyl emphasizes, of course, that the Einstein-Hilbert action is not gauge invariant. Later
work by Pauli \cite{P2} and by Weyl himself \cite{Wey1, Wey2} led soon to the conclusion
that the action (\ref{2.8}) could not be the correct one, and other possibilities were
investigated (see the later editions of Weyl's classic treatise \cite{Wey1}).

Independent of the precise form of the action Weyl shows that in his theory gauge
invariance implies the {\it conservation of electric charge} in much the same way as
general coordinate invariance leads to the conservation of energy and momentum\footnote{We
adopt here the somewhat naive interpretation of energy-momentum conservation for generally
invariant theories of the older literature.}. This beautiful connection pleased him
particularly: ``\ldots [it] seems to me to be the strongest general argument in favour of
the present theory --- insofar as it is permissible to talk of justification in the context
of pure speculation.'' The invariance principles imply five `Bianchi type' identities.
Correspondingly, the five conservation laws follow in two independent ways from the coupled
field equations and may be ``termed the eliminants'' of the latter. These structural
connections hold also in modern gauge theories.

\subsection{Einstein's Objection and Reactions of Other Physicists}

After this sketch of Weyl's theory we come to Einstein's striking counterargument which he
first communicated to Weyl by postcard (see Fig.~3). The problem is that if the idea of a
nonintegrable length connection (scale factor) is correct, then the behavior of clocks
would depend on their history. Consider two identical atomic clocks in adjacent world
points and bring them along different world trajectories which meet again in adjacent world
points. According to (\ref{2.2}) their frequencies would then generally differ. This is in
clear contradiction with empirical evidence, in particular with the existence of stable
atomic spectra. Einstein therefore concludes (see \cite{CPA}, Vol. 8B, Doc. 507):
\begin{quotation}
{\sl \ldots (if) one drops the connection of the $ds$ to the measurement of distance and
time, then relativity looses all its empirical basis.}
\end{quotation}

\begin{figure}\label{fig3}
\epsfig{file=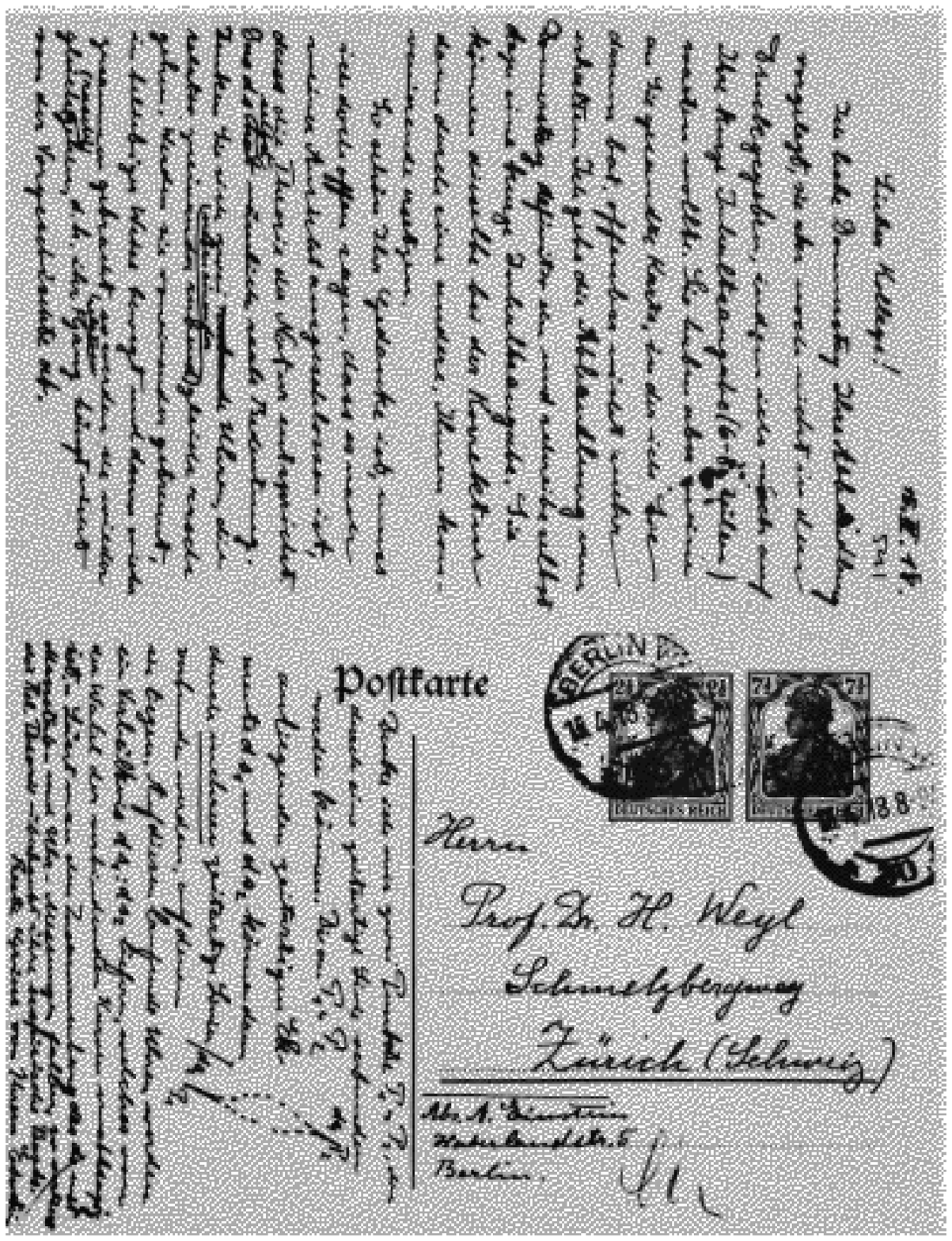,width=12cm} \caption{Postcard of Einstein to Weyl 15.4.1918 (Archives
of ETH).}
\end{figure}

Nernst shared Einstein's objection and demanded on behalf of the Berlin Academy that it
should be printed in a short amendment to Weyl's article. Weyl had to accept this. One of
us has described the intense and instructive subsequent correspondence between Weyl and
Einstein elsewhere \cite{NS1} (see also Vol.~8B of \cite{CPA}). As an example, let us quote
from one of the last letters of Weyl to Einstein (\cite{CPA}, Vol. 8B, Doc. 669):
\begin{quotation}
{\sl This [insistence] irritates me of course, because experience has proven that one can
rely on your intuition; so unconvincing as your counterarguments seem to me, as I have to
admit \ldots}
\end{quotation}

\begin{quotation}
{\sl By the way, you should not believe that I was driven to introduce the linear
differential form in addition to the quadratic one by physical reasons. I wanted, just to
the contrary, to get rid of this `methodological inconsistency {\it (Inkonsequenz)}' which
has been a bone of contention to me already much earlier. And then, to my surprise, I
realized that it looked as if it might explain electricity. You clap your hands above your
head and shout: But physics is not made this way ! (Weyl to Einstein 10.12.1918).}
\end{quotation}

Weyl's reply to Einstein's criticism was, generally speaking, this: The real behavior of
measuring rods and clocks (atoms and atomic systems) in arbitrary electromagnetic and
gravitational fields can be
 deduced only from
a dynamical theory of matter.

Not all leading physicists reacted negatively. Einstein transmitted a very positive first
reaction by Planck, and Sommerfeld wrote enthusiastically to Weyl that there was ``\ldots
hardly doubt, that you are on the correct path and not on the wrong one.''

In his encyclopedia article on relativity \cite{P3} Pauli gave a lucid and precise
presentation of Weyl's theory, but commented on Weyl's point of view very critically. At
the end he states:
\begin{quotation}
{\sl \ldots In summary one may say that Weyl's theory has not yet contributed to getting
closer to the solution of the problem of matter.}
\end{quotation}

Also Eddington's reaction was at first very positive but he soon changed his mind and
denied the physical relevance of Weyl's geometry.

The situation was later appropriately summarized by F. London in his 1927 paper \cite{Lon}
as follows:
\begin{quotation}
{\sl In the face of such elementary experimental evidence, it must have been an unusually
strong metaphysical conviction that prevented Weyl from abandoning the idea that Nature
would have to make use of the beautiful geometrical possibility that was offered. He stuck
to his conviction and evaded discussion of the above-mentioned contradictions through a
rather unclear re-interpretation of the concept of ``real state'', which, however, robbed
his theory of its immediate physical meaning and attraction.}
\end{quotation}
In this remarkable paper, London suggested a reinterpretation of Weyl's principle of gauge
invariance within the new quantum mechanics: The role of the metric is taken over by the
wave function, and the rescaling of the metric has to be replaced by a phase change of the
wave function.

In this context an astonishing early paper by Schr\"odinger~\cite{Sch1} has to be
mentioned, which also used Weyl's ``World Geometry'' and is related to Schr\"odinger's
later invention of wave mechanics. This relation was discovered by V.~Raman and
P.~Forman~\cite{RF}. (See also the discussion by C.N.~Yang in~\cite{Sch2}.)

Even earlier than  London, V.~Fock~\cite{Fo} arrived along a completely different line at
the principle of gauge invariance in the framework of wave mechanics. His approach was
similar to the one by O.~Klein \cite{Kl1}.

The contributions by Schr\"odinger~\cite{Sch1}, London~\cite{Lon} and Fock~\cite{Fo} are
commented in~\cite{LO1}, where also English translations of the original papers can be
found. Here, we concentrate on Weyl's seminal paper ``Electron and Gravitation''.

\section{Weyl's 1929 Classic: ``Electron and Gravitation''}

Shortly before his death late in 1955, Weyl wrote for his {\it Selecta} \cite{Wey4} a
postscript to his early attempt in 1918 to construct a `unified field theory'. There he
expressed his deep attachment to the gauge idea and adds (p.192):
\begin{quotation}
{\sl  Later the quantum-theory introduced the Schr\"odinger-Dirac potential $\psi$ of the
electron-positron field; it carried with it an experimentally-based principle of
gauge-invariance which guaranteed the conservation of charge, and connected the $\psi$ with
the electromagnetic potentials $A_\mu$ in the same way that my speculative theory had
connected the gravitational potentials $g_{\mu\nu}$ with the $A_{\mu}$, and measured the
$A_{\mu}$ in known atomic, rather than unknown cosmological units. I have no doubt but that
the correct context for the principle of gauge-invariance is here and not, as I believed in
1918, in the intertwining of electromagnetism and gravity.}
\end{quotation}

This re-interpretation was developed by Weyl in one of the great papers of this century
\cite{Wey3}. Weyl's classic does not only give a very clear formulation of the gauge
principle, but contains, in addition, several other important concepts and results --- in
particular his two-component spinor theory.

The modern version of the gauge principle is already spelled out in the introduction:

\begin{quotation}
{\sl The Dirac field-equations for $\psi$ together with the Maxwell equations for the four
potentials $f_{p}$ of the electromagnetic field have an invariance property which is
formally similar to the one which I called gauge-invariance in my 1918 theory of
gravitation and electromagnetism; the equations remain invariant when one makes the
simultaneous substitutions
$$\psi\ \ \ {\rm by}\ \ \ e^{i\lambda}\psi\ \ \ \ {\rm and}\ \ \
f_{p}\ \ \ {\rm by}\ \ \  f_{p}-\frac{\partial\lambda}{\partial x^{p}},
$$
where $\lambda$ is understood to be an arbitrary function of position in four-space. Here
the factor $\frac{e}{ch}$, where $-e$ is the charge of the electron, $c$ is the speed of
light, and $\frac{h}{2\pi}$ is the quantum of action, has been absorbed in $f_{p}$. The
connection of this ``gauge invariance'' to the conservation of electric charge remains
untouched. But a fundamental difference, which is important to obtain agreement with
observation, is that the exponent of the factor multiplying $\psi$ is not real but pure
imaginary. $\psi$ now plays the role that Einstein's $ds$ played before. It seems to me
that this new principle of gauge-invariance, which follows not from speculation but from
experiment, tells us that the electromagnetic field is a necessary accompanying phenomenon,
not of gravitation, but of the material wave-field represented by $\psi$. Since
gauge-invariance involves an arbitrary function $\lambda$ it has the character of
``general'' relativity and can naturally only be understood in that context.}
\end{quotation}

We shall soon enter into Weyl's justification which is, not surprisingly, strongly
associated with general relativity. Before this we have to describe his incorporation of
the Dirac theory into GR which he achieved with the help of the tetrad formalism.

One of the reasons for adapting the Dirac theory of the spinning electron to gravitation
had to do with Einstein's recent unified theory which invoked a distant parallelism with
torsion. E.Wigner \cite{Wig} and others had noticed a connection between this theory and
the spin theory of the electron. Weyl did not like this and wanted to dispense with
teleparallelism. In the introduction he says:

\begin{quotation}
{\sl I prefer not to believe in distant parallelism for a number of reasons. First my
mathematical intuition objects to accepting such an artificial geometry; I find it
difficult to understand the force that would keep the local tetrads at different points and
in rotated positions in a rigid relationship. There are, I believe, two important physical
reasons as well. The loosening of the rigid relationship between the tetrads at different
points converts the gauge-factor $e^{i\lambda}$, which remains arbitrary with respect to
$\psi$, from a constant to an arbitrary function of space-time. In other words, only
through the loosening the rigidity does the established gauge-invariance become
understandable. }
\end{quotation}

This thought is carried out in detail after Weyl has set up his two-component theory in
special relativity, including a discussion of $P$ and $T$ invariance. He emphasizes thereby
that the two-component theory excludes a linear implementation of parity and remarks: ``It
is only the fact that the left-right symmetry actually appears in Nature that forces us to
introduce a second pair of $\psi$-components.'' To Weyl the mass-problem is thus not
relevant for this\footnote{At the time it was thought by Weyl, and indeed by all
physicists, that the 2-component theory requires a zero mass. In 1957, after the discovery
of parity nonconservation, it was found that the 2-component theory could be consistent
with a finite mass. See K.M. Case, \protect\cite{Case}.}. Indeed he says: ``Mass, however,
is a gravitational effect; thus there is hope of finding a substitute in the theory of
gravitation that would produce the required corrections.''

\subsection{Tetrad Formalism}


In order to incorporate his two-component spinors into GR, Weyl was forced to make use of
local tetrads (Vierbeine). In section 2 of his paper he develops the tetrad formalism in a
systematic manner. This was presumably independent work, since he does not give any
reference to other authors. It was, however, mainly E.~Cartan who demonstrated with his
work~\cite{Car} the usefulness of locally defined orthonormal bases --also called moving
frames-- for the study of Riemannian geometry.

In the tetrad formalism the metric is described by an arbitrary  basis of orthonormal
vector fields $\{e_{\alpha}(x);\alpha=0,1,2,3\}$. If $\{e^{\alpha}(x)\}$ denotes the dual
basis of 1-forms, the metric is given by
\begin{equation} \label{3.1}
g=\eta_{\mu\nu}e^{\mu}(x)\otimes e^{\nu}(x),\ \ \ \ (\eta_{\mu\nu})=diag(1,-1,-1,-1).
\end{equation}
Weyl emphasizes, of course, that only a class of such local tetrads is determined by
the metric: the metric is not changed if the tetrad fields are subject to
spacetime-dependent Lorentz transformations:

\begin{equation} \label{3.2}
e^{\alpha}(x)\rightarrow\Lambda^{\alpha}_{\ \beta}(x)e^{\beta}(x).
\end{equation}
With respect to a tetrad, the connection forms $\omega=(\omega^{\alpha}_{\ \beta})$ have
values in the Lie algebra of the homogeneous Lorentz group:
\begin{equation} \label{3.3}
\omega_{\alpha\beta}+\omega_{\beta\alpha}=0.
\end{equation}
(Indices are raised and lowered with $\eta^{\alpha\beta}$ and $\eta_{\alpha\beta}$,
respectively.) They are determined (in terms of the tetrad) by the first structure equation
of Cartan:
\begin{equation} \label{3.4}
de^{\alpha}+\omega^{\alpha}_{\ \beta}\wedge e^{\beta}=0.
\end{equation}
(For a textbook derivation see, e.g., \cite{NS2}, especially Sects. 2.6 and 8.5.) Under local Lorentz transformations
(\ref{3.2}) the connection forms transform in the same way as the gauge potential of a
non-Abelian gauge theory:
\begin{equation} \label{3.5}
\omega(x)\rightarrow \Lambda(x)\omega(x)\Lambda^{-1}(x)- d\Lambda(x)\Lambda^{-1}(x).
\end{equation}
The curvature forms $\Omega=(\Omega^{\mu}_{\ \nu})$ are obtained from $\omega$ in exactly
the same way as the Yang-Mills field strength from the gauge potential:
\begin{equation}\label{3.6}
\Omega=d\omega+\omega\wedge\omega
\end{equation}
(second structure equation).

For a vector field $V$, with components $V^{\alpha}$ relative to $\{e_{\alpha}\}$, the
covariant derivative $DV$ is given by
\begin{equation} \label{3.7}
DV^{\alpha}=dV^{\alpha}+\omega^{\alpha}_{\ \beta}V^{\beta}.
\end{equation}
Weyl generalizes this in a unique manner to spinor fields $\psi$:
\begin{equation}\label{3.8}
D\psi=d\psi+\frac{1}{4}\omega_{\alpha\beta}\sigma^{\alpha\beta}\psi.
\end{equation}
Here, the $\sigma^{\alpha\beta}$ describe infinitesimal Lorentz transformations (in the
representation of $\psi$). For a Dirac field these are the familiar matrices
\begin{equation} \label{3.9}
\sigma^{\alpha\beta}=\frac{1}{2}[\gamma^{\alpha},\gamma^{\beta}].
\end{equation}
(For 2-component Weyl fields one has similar expressions in terms of the Pauli matrices.)

With these tools the action principle for the coupled Einstein-Dirac system can be set up.
In the massless case the Lagrangian is
\begin{equation} \label{3.10}
{\cal L}=\frac{1}{16\pi G}R-i\bar{\psi}\gamma^{\mu}D_{\mu}\psi,
\end{equation}
where the first term is just the Einstein-Hilbert Lagrangian (which is linear in $\Omega$).
Weyl discusses, of course, immediately the consequences of the following two symmetries:

(i) local Lorentz invariance,

(ii) general coordinate invariance.

\subsection{The New Form of the Gauge-Principle}

All this is a kind of a preparation for the final section of Weyl's paper, which has the
title ``electric field''. Weyl says:
\begin{quotation}
{\sl We come now to the critical part of the theory. In my opinion the origin and necessity
for the electromagnetic field is in the following. The components $\psi_{1}$ $\psi_{2}$
are, in fact, not uniquely determined by the tetrad but only to the extent that they can
still be multiplied by an arbitrary ``gauge-factor'' $e^{i\lambda}$. The transformation of
the $\psi$ induced by a rotation of the tetrad is determined only up to such a factor. In
special relativity one must regard this gauge-factor as a constant because here we have
only a single point-independent tetrad. Not so in general relativity; every point has its
own tetrad and hence its own arbitrary gauge-factor; because by the removal of the rigid
connection between tetrads at different points  the gauge-factor necessarily becomes an
arbitrary function of position.}
\end{quotation}

In this manner Weyl arrives at the gauge-principle in its modern form and emphasizes:
``From the arbitrariness of the gauge-factor in $\psi$ appears the necessity of introducing
the electromagnetic potential.'' The first term $d\psi$ in (\ref{3.8}) has now to be
replaced by the covariant gauge derivative $(d-ieA)\psi$ and the nonintegrable scale factor
(\ref{2.1}) of the old theory is now replaced by a phase factor:
$$
\exp\left(-\int_{\gamma}A\right)\rightarrow \exp\left(-i\int_{\gamma}A\right),
$$
which corresponds to the replacement of the original gauge group $\mathbb{R}$ by the
compact group $U(1)$. Accordingly, the original Gedankenexperiment of Einstein translates
now to the Aharonov-Bohm effect, as was first pointed out by C.N.~Yang in~\cite{Yan1}. The
close connection between gauge invariance and conservation of charge is again uncovered.
The current conservation follows, as in the original theory, in two independent ways: On
the one hand it is a consequence of the field equations for matter plus gauge invariance,
at the same time, however, also of the field equations for the electromagnetic field plus
gauge invariance. This corresponds to an identity in the coupled system of field equations
which has to exist as a result of gauge invariance. All this is nowadays familiar to
students of physics and does not need to be explained in more detail.

Much of Weyl's paper penetrated also into his classic book ``The Theory of Groups and
Quantum Mechanics'' \cite{Wey5}. There he mentions also the transformation of his early
gauge-theoretic ideas: ``This principle of gauge invariance is quite analogous to that
previously set up by the author, on speculative grounds, in order to arrive at a unified
theory of gravitation and electricity. But I now believe that this gauge invariance does
not tie together electricity and gravitation, but rather electricity and matter.''

When Pauli saw the full version of Weyl's paper he became more friendly and wrote
\cite{PW}:
\begin{quotation}
{\sl In contrast to the nasty things I said, the essential part of my last letter has since
been overtaken, particularly by your paper in Z. f. Physik. For this reason I have
afterward even regretted that I wrote to you. After studying your paper I believe that I
have really understood what you wanted to do (this was not the case in respect of the
little note in the Proc.Nat.Acad.). First let me emphasize that side of the matter
concerning which I am in full agreement with you: your incorporation of spinor theory into
gravitational theory. I am as dissatisfied as you are with distant parallelism and your
proposal to let the tetrads rotate independently at different space-points is a true
solution.}
\end{quotation}

In brackets Pauli adds:
\begin{quotation}
{\sl Here I must admit your ability in Physics. Your earlier theory with $g'_{ik}=\lambda
g_{ik}$ was pure mathematics and unphysical. Einstein was justified in criticizing and
scolding. Now the hour of your revenge has arrived.}
\end{quotation}

Then he remarks in connection with the mass-problem:
\begin{quotation}
{\sl Your method is valid even for the massive {\rm [Dirac]} case. I thereby come to the
other side of the matter, namely the unsolved difficulties of the Dirac theory (two signs
of $m_{0}$) and the question of the 2-component theory. In my opinion these problems will
not be solved by gravitation \ldots the gravitational effects will always be much too
small.}
\end{quotation}

Many years later, Weyl summarized this early tortuous history of gauge theory in an
instructive letter \cite{See} to the Swiss writer and Einstein biographer C.Seelig, which
we reproduce in
an English translation.
\begin{quotation}
{\sl The first attempt to develop a unified field theory of gravitation and
electromagnetism dates to my first attempt in 1918, in which I added the principle of
gauge-invariance to that of coordinate invariance. I myself have long since abandoned this
theory in favour of its correct interpretation: gauge-invariance as a principle that
connects electromagnetism not with gravitation but with the wave-field of the electron.
---Einstein was against it} [the original theory] {\sl from the beginning, and this led to
many discussions. I thought that I could answer his concrete objections. In the end he said
``Well, Weyl, let us leave it at that! In such a speculative manner, without any guiding
physical principle, one cannot make Physics.'' Today one could say that in this respect we
have exchanged our points of view. Einstein believes that in this field} [Gravitation and
Electromagnetism] {\sl the gap between ideas and experience is so wide that only the path
of mathematical speculation, whose consequences must, of course, be developed and
confronted with experiment, has a chance of success. Meanwhile my own confidence in pure
speculation has diminished, and I see a need for a closer connection with quantum-physics
experiments, since in my opinion it is not sufficient to unify Electromagnetism and
Gravity. The wave-fields of the electron and whatever other irreducible elementary
particles may appear must also be included. }

Independently of Weyl, V.~Fock~\cite{Fo2} also incorporated the Dirac equation into GR by
using the same method. On the other hand, H.~Tetrode~\cite{Tet},
E.~Schr\"odinger~\cite{Sch3} and V.~Bargmann~\cite{Bar} reached this goal by starting with
space-time dependent $\gamma$-matrices, satisfying
$\left\lbrace\gamma^{\mu},\,\gamma^{\nu}\right\rbrace=2\,g^{\mu\nu}$. A somewhat later work
by L.~Infeld and B.L.~van der Waerden~\cite{Wae} is based on spinor analysis.
\end{quotation}

\section{Concluding Remarks}

Gauge invariance became a serious problem when Heisenberg and Pauli began to work on a
relativistically invariant QED that eventually resulted in two important papers ``On the
Quantum Dynamics of Wave Fields'' \cite{HP1}, \cite{HP2}. Straightforward application of
the canonical formalism led, already for the free electromagnetic field, to nonsensical
results. Jordan and Pauli on the other hand, proceeded to show how to quantize the theory
of the \emph{free field} case by dealing only with the field strengths $F_{\mu\nu}(x)$. For
these they found commutation relations at different space-time points in terms of the now
famous invariant Jordan-Pauli distribution that are manifestly Lorentz invariant.

The difficulties concerned with applying the canonical formalism to the electromagnetic
field continued to plaque Heisenberg and Pauli for quite some time. By mid-1928 both were
very pessimistic, and Heisenberg began to work on ferromagnetism\footnote{Pauli turned to literature.
In a letter of 18 February 1929 he wrote from Z\"{u}rich to Oskar Klein:
``For my proper amusement I then made a
short sketch of a utopian novel which was supposed to have the
title `Gulivers journey to Urania' and was intended as a political
satire in the style of Swift against present-day democracy. [...]
Caught in such dreams, suddenly in January, news from Heisenberg
reached me that he is able, with the aid of a trick ... to get rid
of the formal difficulties that stood against the execution of our
quantum electrodynamics.'' \cite{P1}}. In fall of 1928
Heisenberg discovered a way to bypass the difficulties. He added the term
$-\frac{1}{2}\varepsilon(\partial_\mu A^\mu)^2$ to the Lagrangian, in which case the component
$\pi_0$ of the canonical momenta
\[\pi_\mu=\frac{\partial L}{\partial_0A_\mu}\]
does no more vanish identically ($\pi_0=-\varepsilon \partial_\mu A^\mu$). The standard canonical
quantization scheme can then be applied. At the end of all calculations one could then take the limit
$\varepsilon\rightarrow0$.

In their second paper, Heisenberg and Pauli stressed that the Lorentz condition cannot be imposed
as an operator identity but only as a supplementary condition selecting admissible states. This
discussion was strongly influenced by a paper of Fermi from May 1929.

For this and the further main developments during the early period of quantum field theory, we refer
to chapter 1 of \cite{SS}.

As in Weyl's work GR also played a crucial role in Pauli's discovery of non-Abelian gauge
theories. (See Pauli's letters to Pais and Yang in  Vol. 4 of \cite{P1}). He arrived at all
basic equations through dimensional reduction of a generalization of Kaluza-Klein theory,
in which the internal space becomes a two-sphere. (For a description in modern language,
see \cite{RS}).

In contrast, in the work of Yang and Mills  GR played no role. In an interview Yang said on
this in 1991:
\begin{quote}
``\textit{It happened that one semester [around 1970] I was teaching GR, and I noticed that
the formula in gauge theory for the field strength and the formula in Riemannian geometry
for the Riemann tensor are not just similar -- they are, in fact, the same if one makes the
right identification of symbols! It is hard to describe the thrill I felt at understanding
this point.}''
\end{quote}

The developments after 1958 consisted in the gradual recognition that---contrary to
phenomenological appearances---Yang-Mills gauge theory could describe weak and strong
interactions. This important step was again very difficult, with many hurdles to overcome.

\end{document}